\documentclass[preprint]{aastex}

\pdfoutput=1

\usepackage{graphicx}
\usepackage{times}

\shorttitle{Optical degradation on the moon}
\shortauthors{Murphy et al.}

\begin{document} 

\title{Long-term degradation of optical devices on the moon}

\author{
T.\,W. Murphy,~Jr.\altaffilmark{1},
E.\,G. Adelberger\altaffilmark{2},
J.\,B.\,R. Battat\altaffilmark{3},
C.\,D. Hoyle\altaffilmark{4},
R.\,J. McMillan\altaffilmark{5},
E.\,L. Michelsen\altaffilmark{1},
R.\,L. Samad\altaffilmark{1},
C.\,W. Stubbs\altaffilmark{6},
H.\,E. Swanson\altaffilmark{2}}
\email{tmurphy@physics.ucsd.edu}

\altaffiltext{1}{University of California, San Diego, Dept. of Physics, La Jolla, CA 92093-0424}
\altaffiltext{2}{University Washington, Dept. of Physics, Seattle, WA 98195-1560}
\altaffiltext{3}{Massachusetts Institute of Technology, Dept. of Physics, Cambridge, MA 02139}
\altaffiltext{4}{Humboldt State University, Dept. of Physics and Astronomy, Arcata, CA 95521-8299}
\altaffiltext{5}{Apache Point Observatory, Sunspot, NM 88349-0059}
\altaffiltext{6}{Harvard University, Dept. of Physics, Cambridge, MA 02318}

\begin{abstract}

Forty years ago, Apollo astronauts placed the first of several
retroreflector arrays on the lunar surface.  Their continued usefulness for
laser-ranging might suggest that the lunar environment does not
damage optical devices.  However, new laser ranging data reveal that the
efficiency of the three Apollo reflector arrays is now diminished by a factor
of ten at all lunar phases and by an additional factor of ten when the
lunar phase is near full moon.  These deficits did not exist in the
earliest years of lunar ranging, indicating that the lunar environment
damages optical equipment on the timescale of decades.  Dust or abrasion on
the front faces of the corner-cube prisms may be responsible, reducing
their reflectivity and degrading their thermal performance when exposed to
face-on sunlight at full moon.  These mechanisms can be tested using
laboratory simulations and must be understood before designing equipment
destined for the moon.

\end{abstract}


\section{Introduction}

Long-term NASA plans \citep{nasa-plan} for placing scientific equipment on
the moon face uncertainty regarding the environmental impact on such
devices as hard information about the lunar environmental effect on
scientific instruments has not been available.  From a quantitative
analysis of the performance of the laser reflectors, we find clear evidence for
degradation of the retroreflectors, and note that 
degradation began within one decade of placement on the lunar surface.

From 1969--1985, the McDonald Observatory 2.7~m Smith Telescope \citep[MST:][]{bender}  
dominated lunar laser ranging (LLR), using a 634~nm ruby laser. Starting
around 1985, the McDonald operation moved away from the competitively-scheduled
MST to a dedicated 0.76~m telescope designed to perform
both satellite and lunar laser ranging, becoming the McDonald Laser
Ranging System \citep[MLRS:][]{mlrs}.  In 1984, other LLR operations
began at the Observatoire de la C\^ote d'Azur \citep[OCA:][]{oca} in France
and at the Haleakala site in Hawaii, that used 1.5~m
and 1.74~m telescopes, respectively.  These systems all operate Nd:YAG
lasers at 532~nm.

In 2006, the Apache Point Observatory Lunar Laser-ranging Operation \citep[APOLLO:][]{apollo}
began science operations using the 3.5~m telescope and a 532~nm laser at the
Apache Point Observatory in New Mexico. Primarily geared
toward improving tests of gravity, APOLLO is designed to reach a range
precision of one millimeter via a substantial increase in the rate
of return photons. The large telescope aperture and good image quality
at the site, when coupled with a $4\times4$ single-photon detector
array, produces return photon rates 
from all three Apollo reflectors that are about 70 times higher than
the best rates experienced by the previous LLR record-holder (OCA).
Consequently, APOLLO is able to obtain ranges through the full moon
phase for the first time since MST
LLR measurements ceased around 1985. 

We find that the performance of the reflectors themselves degrades during
the period surrounding full moon.  In this paper we describe the full-moon
deficit, report its statistical significance, and eliminate the possibility
that it results from reduced system sensitivity at full moon. We show
that this deficit began in the 1970's, and examine the
significance of successful total-eclipse observations by OCA and MLRS.  We
see an additional factor-of-ten signal deficit that applies at all lunar
phases, but this observation requires a detailed technical evaluation of
the link, and is deferred to a later publication.  We
briefly discuss possible mechanisms that might account for the observed
deficits.

\section{Degradation at full moon}

APOLLO observing sessions typically last less than one hour, with
a cadence of one observing session every 2--3 nights. For a variety
of practical reasons, APOLLO observations are confined to 75\% of
the lunar phase distribution, from $D=45^{\circ}$ to $D=315^{\circ}$,
where $D$ is the synodic phase relative to new moon at $D=0$. Within
an observing session, multiple short ``runs'' are carried out, where
a run is defined as a contiguous sequence of laser shots to a specific
reflector. Typical runs last 250 or 500 seconds, consisting of 5000
or 10000 shots at a 20~Hz repetition rate. Each shot sends about
$10^{17}$ photons toward the moon, and in good conditions 
we detect about one return photon per shot. If the signal level
acquired on the larger Apollo~15 reflector is adequate, we cycle to the
other two Apollo reflectors in turn, sometimes completing multiple
cycles among the reflectors in the allotted time. When the Lunokhod~2
reflector is in the dark, we range to it as well. Its design leads
to substantial signal degradation from thermal gradients, rendering
it effectively unobservable during the lunar day.

\begin{figure}[tb]
\begin{center}\includegraphics[width=89mm]{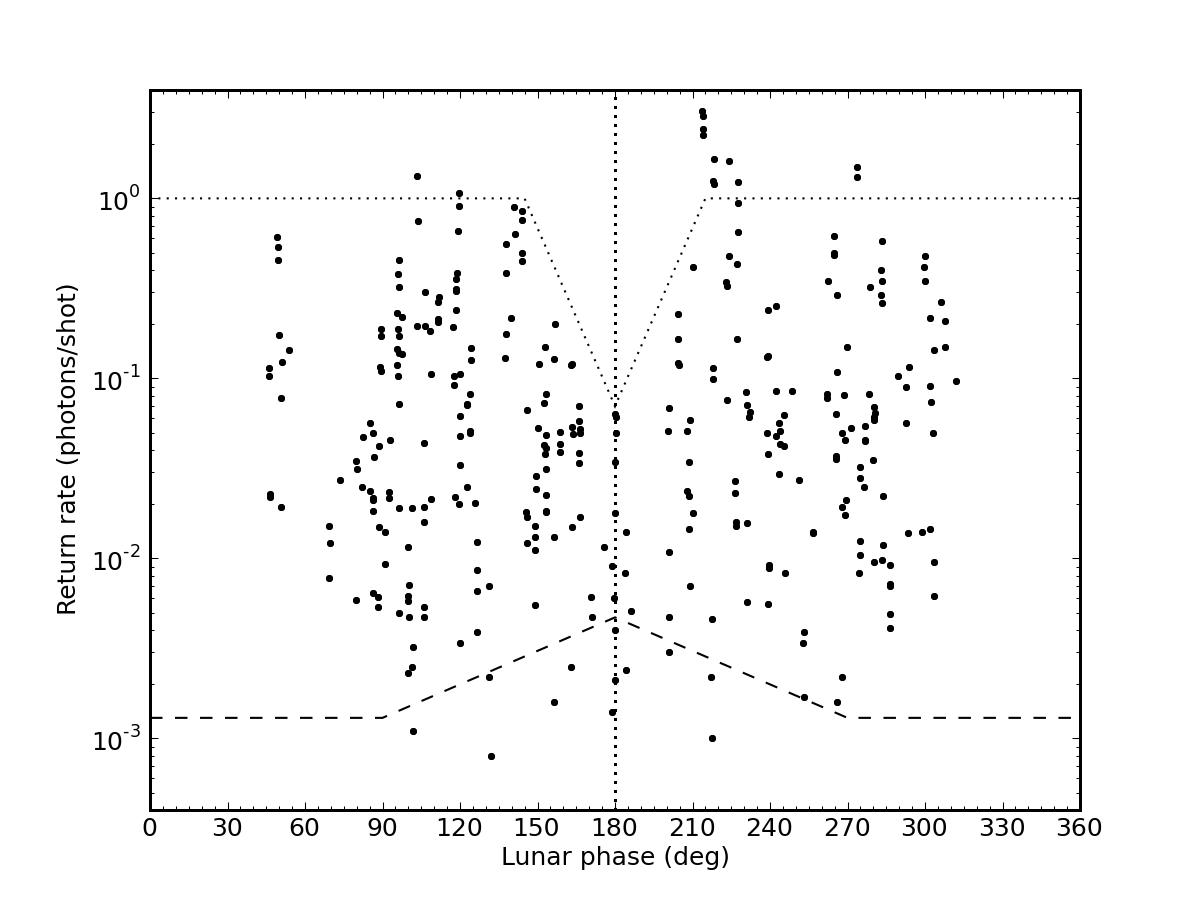}
\end{center}
\caption{APOLLO return rate (in photons per shot) from the Apollo~15 reflector
as a function of lunar phase. The reduced signal strength around full moon
($D=180^{\circ}$) is apparent. The vertical scatter is predominantly
due to variable atmospheric seeing and transparency. The dotted line
across the top is a simple ad-hoc model of the signal deficit used
to constrain background suppression in Fig.~\ref{fig:background}.
The dotted line across the bottom indicates the background rate in
a 1~ns temporal window, against which signal identification must
compete.\label{fig:APOLLO-rate}}
\end{figure}

Figure~\ref{fig:APOLLO-rate} displays APOLLO's return rates, in return
photons per shot, for the Apollo~15 reflector as a function of lunar phase,
with 338 data points spanning 2006-10-03 to 2009-06-15.  Signal rate is
highly dependent on atmospheric seeing (turbulence-related image quality).
When the seeing is greater than 2~arcsec, the signal rate scales like the
inverse fourth power of the seeing scale \citep{apollo}. Variability in
seeing and transparency dominate the observed spread of signal strength,
resulting in at least two orders-of-magnitude of variation.

Below about 0.001 photons per shot (pps), we have difficulty identifying
the signal against photon background and detector dark rate. A typical
peak rate across phases is $\sim1$~pps, with the best runs reaching
$\sim3$~pps. \emph{The key observation is the order-of-magnitude
dip in signal rate in the vicinity of full moon}, at $D=180^{\circ}$.
The best return rates at full moon were associated with pristine observing
conditions that would have been expected to deliver $\sim1$ photon
per shot at other phases, but only delivered 0.063~pps at full moon.
Thus the deficit is approximately a factor of 15. The deficit appears
to be confined to a relatively narrow range of $\pm30^{\circ}$ around
the full moon, and is not due to uncharacteristically poor observing
conditions during this period.

A Kolmogorov-Smirnov (K-S) test confirms the improbability that random
chance could produce a full-moon dip as large as that seen in
Fig.~\ref{fig:APOLLO-rate}.  There is $<0.03$\% chance that the
measurements within $\pm 30^\circ$ of full moon were drawn from the same
distribution as the out-of-window points.   Similar tests using 
$60^\circ$-wide windows centered away from full moon do not produce
comparably low probabilities.

\begin{figure}[tb]
\begin{center}\includegraphics[width=89mm]{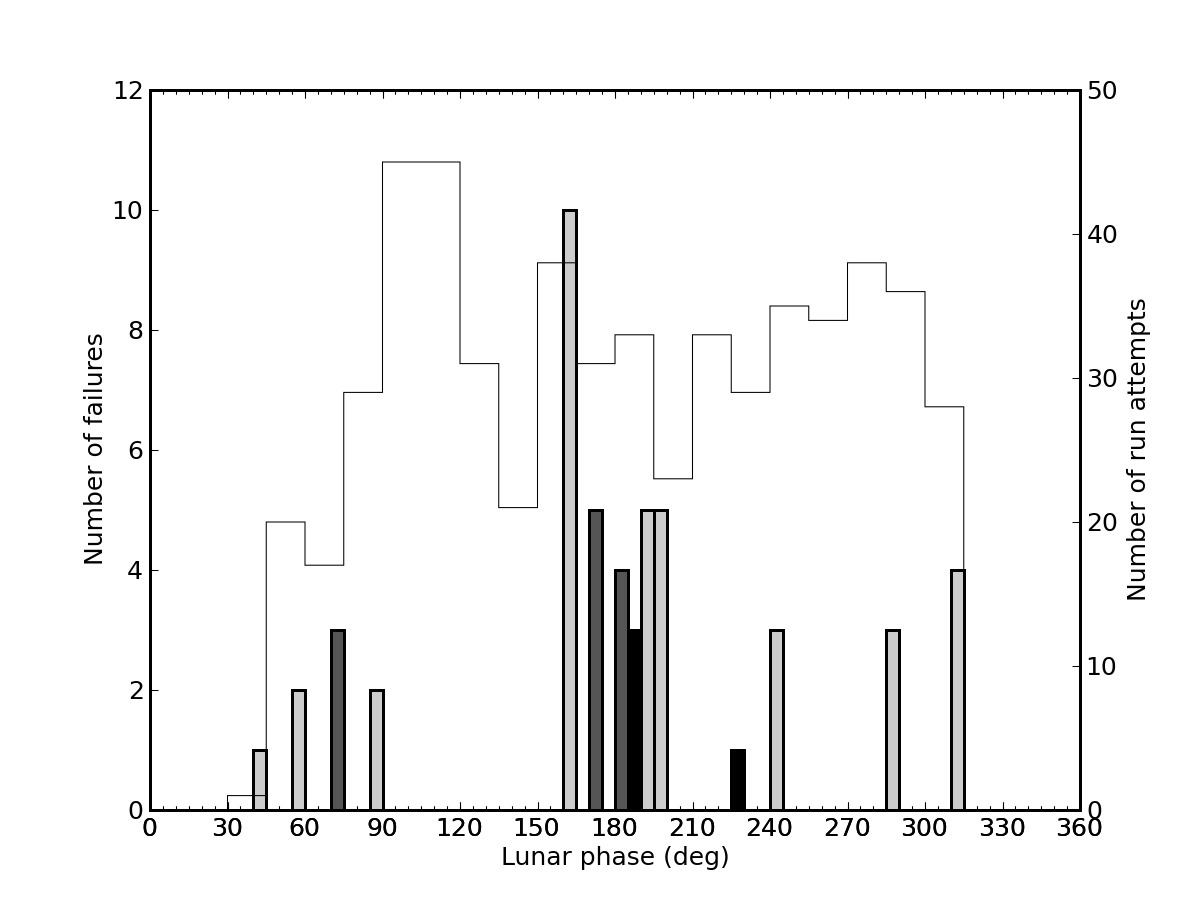}
\end{center}
\caption{Phase distribution in 5$^\circ$ bins of failed run attempts (left-hand scale) on Apollo~15
during the period from 2006-10-03 to 2009-06-15, excluding those due
to known technical difficulties. Black indicates good observing conditions,
medium gray corresponds to medium conditions, and light gray reflects
bad conditions. The line histogram shows the phase distribution of
all run attempts in 15$^\circ$ bins (right-hand scale).\label{fig:failure}}
\end{figure}

Additional evidence for the full-moon
deficit is provided by the instances of failure
to acquire a signal. Failure can occur for a variety of reasons \emph{not}
related to the health of the lunar arrays: poor seeing; poor atmospheric
transparency; inaccurate telescope pointing; optical misalignment
between transmit and receive beams; time-of-flight prediction error;
instrumental component failure. But none of these causes depend on
the phase of the moon. We therefore plot a histogram of Apollo~15
acquisition failures as a function of lunar phase in Fig.~\ref{fig:failure}.
Failures due to known instrumental problems were removed from
this analysis, as were failures due to causes such as pointing
errors that were ultimately remedied within the session. The bars
are shaded to reflect observing conditions: light gray indicates 
poor conditions (seeing or transparency); medium gray
indicates medium conditions; and black indicates excellent observing
conditions, for which the lack of signal is especially puzzling. Note
the cluster of failures centered around full moon. The phase distribution
of run attempts is roughly uniform.

The other Apollo reflectors are similarly impacted at full moon. On
the few occasions when the full-moon Apollo~15 signal was strong
enough to encourage attempts on the other reflectors, we found that
the expected 1:1:3 ratio between the Apollo 11, 14, and 15 rates is
approximately preserved. In no case have we been able to raise a signal
on other reflectors after repeated failures to acquire signal from
Apollo~15.

\begin{figure}[tb]
\begin{center}\includegraphics[width=89mm]{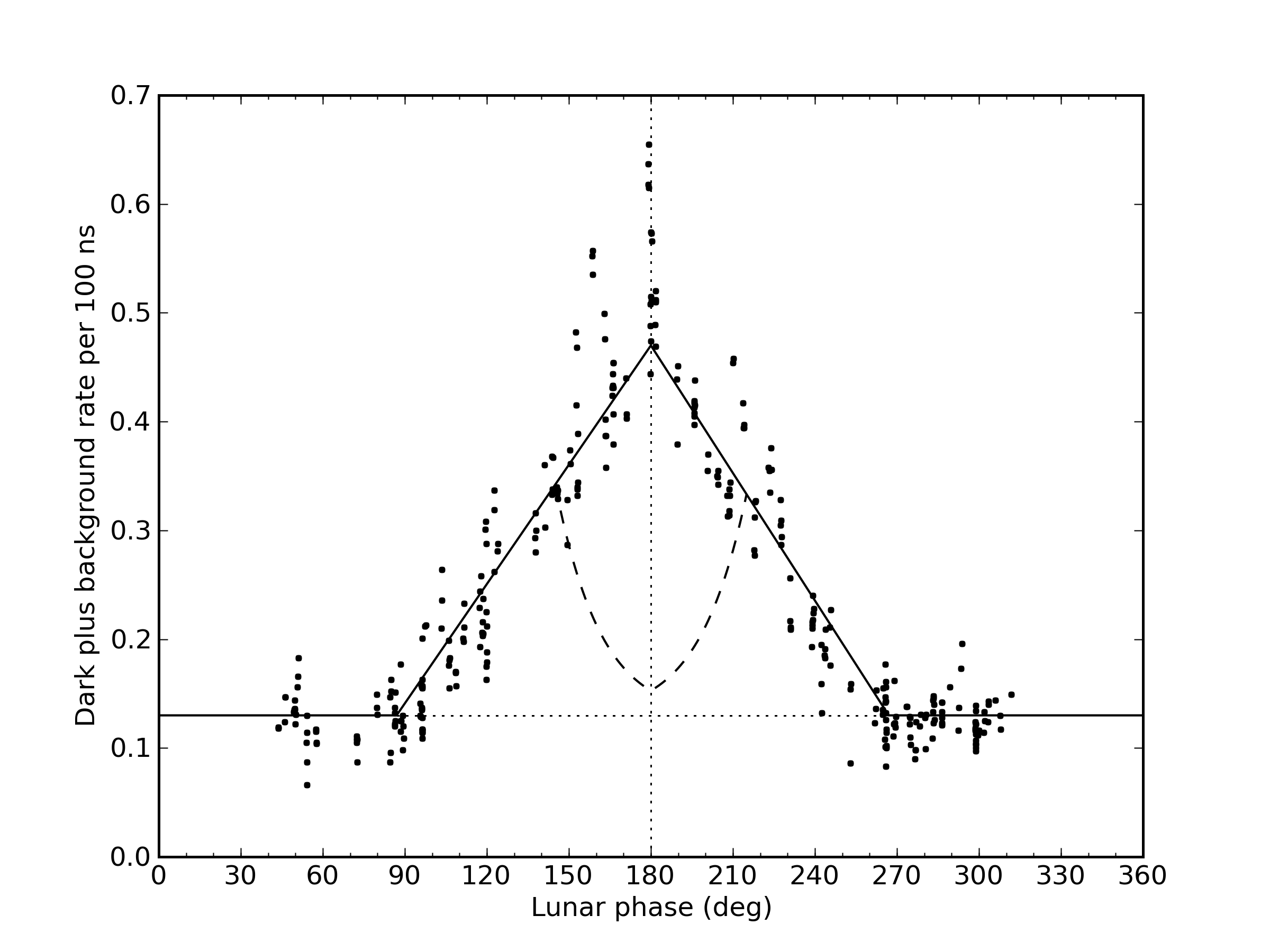}
\end{center}
\caption{APOLLO's background rate from the Apollo~15 site. The
dotted line shows the dark-rate baseline and the solid line the contribution
of the lunar illumination (not including the extra 40\% enhancement
at full moon). The dashed curve shows the expected background rate
if the detector's full-moon sensitivity were suppressed in
the same way as the lunar signal seen in Fig.~\ref{fig:APOLLO-rate}.
APOLLO's detector clearly has high sensitivity
at full moon.\label{fig:background}}
\end{figure}

Could the full-moon deficit be explained by paralysis of our single-photon
avalanche photodiode (APD) detectors in response to the increased
background at full moon? Figure~\ref{fig:background} indicates that APOLLO
sees a maximum background rate at full moon of $\sim0.6$ avalanche events
per 100~ns detection gate across the $4\times4$ APD array---in agreement
with throughput calculations. Therefore, a typical gate-opening has a
$\sim$30\% chance that \emph{one} of the 11 consistently-functioning
avalanche photodiode elements (out of 16) will be rendered blind
\emph{prior} to the arrival of a lunar photon halfway into the 100~ns gate.
The sensitivity for the entire array to signal return photons therefore
remains above 97\% even at full moon. The background rates presented here
are extrapolated from a 20~ns window in the early part of the gate, before
any lunar return signal.

The Apollo~15 site is near the lunar prime meridian, so that its
illumination curve is roughly symmetric about full moon.  Small-aperture
photometry measurements by \citet{peacock}---and more recently by
\citet{kieffer}---show that the surface brightness
increases roughly linearly on approach to full moon, with an additional $\sim40$\%
enhancement very near full moon \citep{opposition}. A linear illumination
curve is provided in Fig.~\ref{fig:background} for reference. APOLLO
clearly sees the expected background enhancement at full moon. If
\emph{any} phenomenon suppressed APD sensitivity to laser returns from the
reflector array, it would \emph{likewise} suppress sensitivity to the
background photons, 
as suggested by the dashed curve in
Fig.~\ref{fig:background}.  There is no hint of detector suppression in the
background counts, so we conclude that the diminished return rate observed
near full moon constitutes a genuine reduction in signal returning from the
reflector.

\begin{figure}[tb]
\begin{center}\includegraphics[width=165mm]{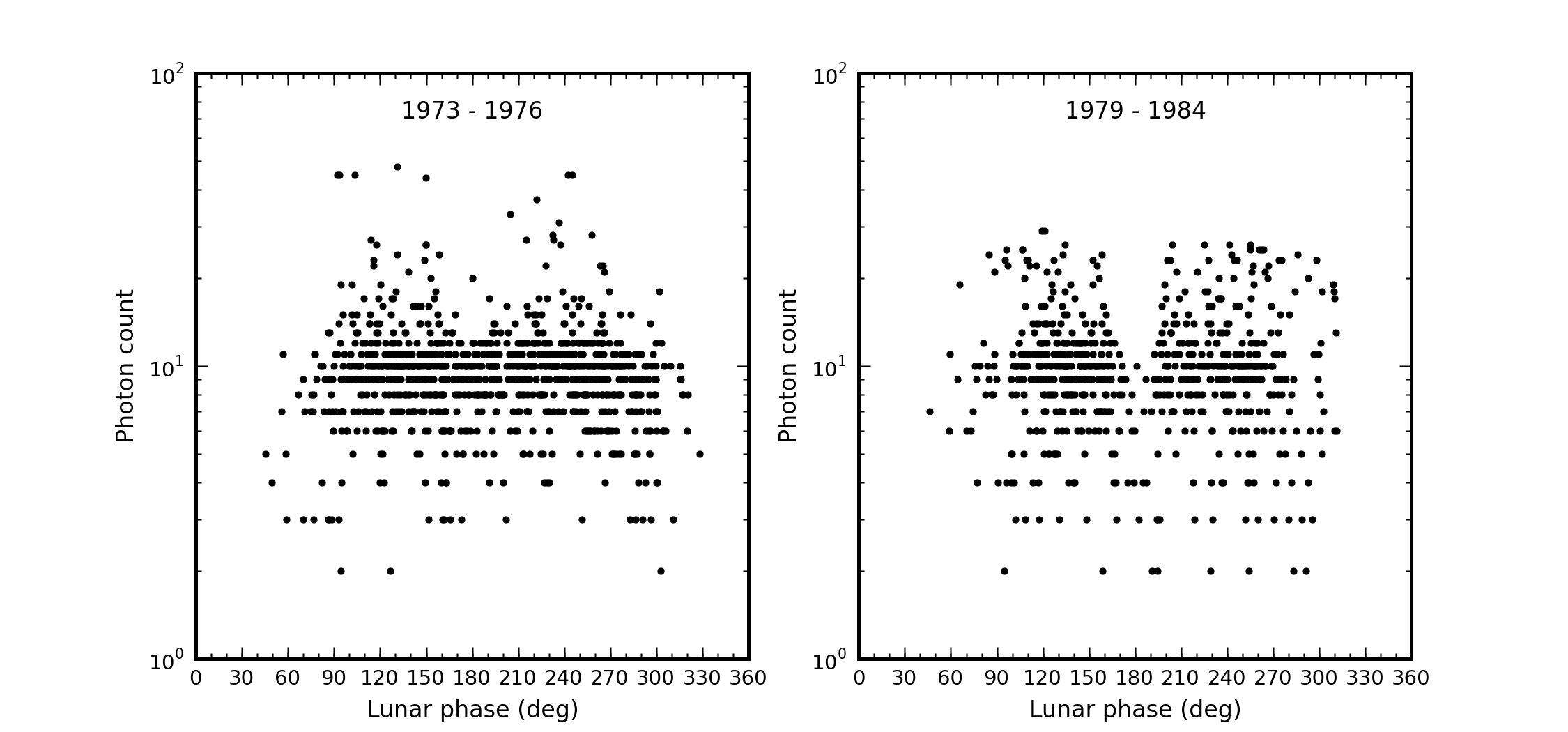}
\end{center}
\caption{Apollo~15 photon counts detected at MST in
the four years spanning 1973--1976 and six years spanning 
1979--1984 \citep{cddis}. The first period lacks any convincing drop in signal rate
around full moon, while the later period reveals the emergence of
a full-moon deficit. The probability that the full-moon distribution
is compatible with the rest of the points is 18\% and 0.03\% for the
two periods, respectively. The vertical spread of these points is smaller
than the APOLLO spread in
Fig.~\ref{fig:APOLLO-rate} because the MST's
larger beam divergence and receiver aperture made it less
sensitive to atmospheric seeing. \label{fig:old_mcd}}
\end{figure}

It is natural to ask if we can determine the timescale over which the
full-moon deficit developed. MST LLR data \citep{cddis,ilrs} reveal that from
1973 to 1976 there was no indication of a full-moon deficit.
Figure~\ref{fig:old_mcd} shows the photon count per run for two periods of
the MST operation, where a run typically consisted of 150--200 shots at a
rate of 20 shots per minute. A full-moon deficit began to develop in the
period from 1977--1978 (not shown), and is markedly evident in the period
from 1979--1984---but it appears somewhat narrower than the deficit now
observed by APOLLO. K-S tests indicate that the probability that the
distribution of photon counts between $160^{\circ}<D<200^{\circ}$ is the
same as that outside the window is 18\%, 0.9\%, and 0.03\% for the three
periods indicated above, in time-order. In the last period, roughly one
decade after placement of the Apollo~15 array in 1971, the deficit was
approximately a factor of three.  The MST apparatus did not change in a
substantial way between 1973 to 1984.

By 1985, LLR was performed only on smaller telescopes, and attempts at full
moon ranging subsided---except during four total lunar eclipses.  The 35
eclipse range measurements by OCA and MLRS add an interesting twist: the
return strength during eclipse is statistically indistinguishable from that
at other phases and not compatible with an order-of-magnitude signal
deficit. Existing data do not probe the time evolution of reflector
efficiency into and out of the total eclipse, but it appears that the
arrays perform normally as soon as 15 minutes into the totality. APOLLO
could not observe the eclipses of 2007 August 28 and 2008 February 21
because of bad weather, but will have a chance to follow a complete total
eclipse on 2010 December 21.

\section{Other evidence for degradation}

In addition to the full-moon deficit, analysis of APOLLO's return rate
reveals an overall factor-of-ten signal deficit at all lunar phases.
Supporting evidence requires more analysis than can be covered here. In
brief, the dominant contributors to the $\sim10^{17}$ photon throughput
loss arise from beam divergence on both the uplink and downlink. We can
measure the former by deliberately scanning the beam across the reflector
on the moon, confirming a seeing-limited beam profile. We additionally
measure the atmospheric seeing via the spatial distribution of the return
point source on the $4\times4$ detector array.  The downlink divergence is
set by diffraction from the corner cubes, verified by measurements of the
actual flight cubes. Receiver throughput losses, which constitute a small
fraction of the total loss, were measured by imaging stars or the bright
lunar surface on the APD, and agree well with a model of the optical and
detector system. Careful analysis does not account for APOLLO's missing
factor of ten in signal return, while early ranging data from MST do agree
with the anticipated return rate \citep{llr-1970}.

Further evidence for the damaging effects of the lunar environment comes
from the Lunokhod~2 reflector.  In the first six months of Lunokhod~2
observations in 1973, its signal was 25\% stronger than that from the
Apollo~15 array. Today, we find that it is 10 times weaker.
The Lunokhod corner cubes are more exposed than the recessed Apollo cubes,
and unlike the Apollo cubes, have a silver coating on the rear surfaces.
Both factors may contribute to the accelerated degradation of the 
Lunokhod array relative to the Apollo arrays.

\section{Discussion}

The full-moon deficit, the overall signal shortfall experienced by APOLLO,
and the relative decline in performance of the Lunokhod array all show that
the lunar reflectors have degraded with time.  It may be possible to
explain these observations with a single mechanism that causes both an
optical impairment at all phases, and a thermal influence near full moon
that abates during eclipse.  One possibility is alteration of the
corner-cube prisms' front surfaces either via dust deposition or surface
abrasion from high-velocity impact ejecta or micrometeorites.
Alternatively, any material coating on the back of the corner
cubes---perhaps originating from the teflon mounting rings---could impact
performance of the Apollo reflectors via frustration of total internal
reflection (TIR) and absorption of solar energy.  Bulk absorption in the
glass could also produce the observed effects.

The impact on reflection efficiency at all phases from each of these
possibilities is obvious.  The full-moon effect would arise from an
enhancement of solar energy absorption by the corner-cube prisms and their
housings---defeating the careful thermal design intended to keep the prisms
nearly isothermal.  Because the uncoated Apollo corner cubes work via TIR,
their rejection of solar flux should be complete when sunlight arrives
within 17$^\circ$ of normal incidence.  But the temperature uniformity
within the corner cubes is upset either by absorption of energy at the cube
surfaces, or by defeat of TIR via scattering---which results in energy
deposition in the pocket behind the cubes, heating the cubes from the rear.
Temperature gradients in a corner-cube prism produce refractive index
gradients, generating wavefront distortion within the prism.  A 4~K
gradient between the vertex and front face of the Apollo corner cubes
reduces the peak intensity in their far-field diffraction pattern by a
factor of ten \citep[][Fig. 10]{adl}.  Apollo corner cubes are recessed by
half their diameter in a tray oriented toward the earth. Near full moon,
the weathered corner cubes are most fully exposed to solar illumination,
maximizing the degradation.  During eclipses, the reflector response may be
expected to recover on a short timescale, governed by the $\sim15$~minute
thermal diffusion timescale for 38~mm diameter fused silica corner-cube
prisms.

While any of the proposed mechanisms could account for the observations,
objections can be raised to each of them.  Bulk absorption is not expected
in the Suprasil fused silica used for the Apollo cubes after 40 years of
exposure.  Micrometeorite rates on the lunar surface gleaned from study of
return samples, and summarized in \citet{surveyor}, suggest the fill-factor
of craters on an exposed surface to be $\sim 10^{-4}$ after 40 years,
dominated by craters in the 10--100~$\mu$m range.  Opportunities for a
contaminant coating on the rear surfaces of the corner cubes are limited
given that the only substance within the closed aluminum pocket besides the
glass corner cube is the teflon support ring.  Moreover, the Lunokhod array
would not be subject to the same rear-surface phenomena as the Apollo
cubes, yet shows an even more marked degradation.

Dust is perhaps the most likely candidate for the observed degradation.
Astronaut accounts from the surface and from lunar orbit, as well as a
horizon glow seen by Surveyor~7, suggest the presence of levitated
dust---possibly to altitudes in excess of 100~km, for which a lofting
mechanism has been suggested by \citet{stubbs-dust}.  The dust monitor
placed on the lunar surface by the Apollo~17 mission measured large fluxes
of dust in the east-west direction around the time of lunar sunrise and
sunset---consistent with the electrostatic charging mechanisms described by
\citet{farrell-terminator}.  The main difficulty with the dust explanation
is that electrostatic charging alone is not strong enough to liberate dust
grains from surface adhesion.  But mechanical disturbance seeded by
micrometeorite and impact ejecta activity may be enough to free the
already-charged grains.  Whether or not dust is responsible, the supposed
health of the reflector arrays has been used to argue that dust dynamics on
the surface of the moon are of minimal importance.  Our observations of the
reduced reflector performance invalidate the invocation of reflector health in
this argument.

The only other relevant data for the environmental impact on optical devices
on the lunar surface comes from the Surveyor~3 camera lens, retrieved by
the Apollo~12 mission.  After 945 days on the surface, the glass cover of
the camera lens had dust obscuring an estimated 25\% of its
surface---though it is suspected that much of this was due to Surveyor and
Apollo~12 landing and surface activities \citep{surveyor}.  Clearly, the
ascent of the lunar modules could result in dust deposition on the nearby
reflectors.  But the effect reported here became established
after several years on the lunar surface (e.g., Fig.~\ref{fig:old_mcd}),
and is therefore not related to liftoff of the lunar modules.

The evidence for substantially worsened performance of the lunar reflectors
over time makes it important to consider the long-term usefulness of
next-generation devices proposed for the lunar surface. Finding the
mechanism responsible for the observed deficits is a high priority.
Thermal simulations or testing deliberately altered corner-cube prisms in a
simulated lunar environment would likely expose the nature of the problem
with the Apollo arrays.  Especially important would be to differentiate
between permanent abrasion versus removable dust.  The results could impact
the designs of a wide variety of space hardware---especially
next-generation laser ranging reflectors, telescopes, optical communication
devices, or equipment dependent on passive thermal control.

\acknowledgments

We thank Doug Currie, Eric Silverberg, and Kim Griest for comments.  APOLLO
is indebted to the staff at the Apache Point Observatory, and to Suzanne
Hawley and the University of Washington Astronomy Department for APOLLO's
telescope time. We also acknowledge the technological prowess of Apollo-era
scientists and engineers, who designed, tested, and delivered the first
functional reflector array for launch on Apollo~11 within six months of
receipt of the contract. APOLLO is jointly funded by NSF and NASA, and some
of this analysis was supported by the NASA Lunar Science Institute as part
of the LUNAR consortium (NNA09DB30A).


\begin{thebibliography}{1}

\bibitem[Alley et al. (1970)]{llr-1970}Alley, C.\,O., Chang, R.\,F., Currie, D.\,G. \& Poultney,
S.\,K., 19870.  Laser ranging retro-reflector: Continuing measurements and expected
results. \emph{Science} \textbf{167}, 458--460

\bibitem[Arthur D. Little (1969)]{adl}Arthur D. Little, Inc. 1969.  \emph{Laser Ranging Retro-Reflector Array
for the Early Apollo Scientific Experiments Package}, available at:
\url{www.physics.ucsd.edu/\~{}tmurphy/apollo/doc/ADL.pdf}

\bibitem[Bender et al. (1973)]{bender}Bender, P.\,L., and 12 colleagues, 1973.  The lunar laser ranging experiment.
\emph{Science} \textbf{182}, 229--238

\bibitem[CDDIS (2009)]{cddis}CDDIS, 2009. Data available at
\url{ftp://cddis.gsfc.nasa.gov/pub/slr/data/npt/moon/} 

\bibitem[Farrell et al. (2007)]{farrell-terminator}Farrell, W.\,M., Stubbs,
T.\,J., Vondrak, R.\,R., Delory, G.\,T., \& Halekas, J.\,S., 2007.  Complex
electric fields near the lunar terminator: The near-surface wake and
accelerated dust. \emph{Geophysical Research Letters}, \textbf{34}, L14201
(5 pages)

\bibitem[Hapke, Nelson, \& Smythe (1998)]{opposition}Hapke, B., Nelson, R., \& Smythe, W., 1998.  The opposition
effect of the moon: Coherent backscatter and shadow hiding. \emph{Icarus}
\textbf{133}, 89--97

\bibitem[Johnson, Taylor, \& Wetzel (1992)]{surveyor}Johnson, S.\,W.,
Taylor, G.\,J., \& Wetzel, J.\,P., 1992.  Environmental effects on lunar
astronomical observatories. \emph{Lunar Bases and Space Activities of the
21$^{\rm st}$ Century II}, ed. Mendell, W.\,W. et al., 329--335

\bibitem[Kieffer \& Stone (2005)]{kieffer}Kieffer, H.\,H. \& Stone, T.\,C.,
2005.  The spectral irradiance of the moon.  \emph{Astron. J.}, \textbf{129},
2887--2901

\bibitem[Murphy et al. (2008)]{apollo}Murphy, T.\,W. Jr. et al., and 12 colleagues, The Apache Point Observatory
lunar laser-ranging operation: Instrument description and first detections.
\emph{Publ. Astron. Soc. Pac}. \textbf{120}, 20--37

\bibitem[NASA Strategic Plan (2006)]{nasa-plan}NASA, 2006. \emph{2006 NASA Strategic Plan},
\url{www.nasa.gov/pdf/142302main\_2006\_NASA\_Strategic\_Plan.pdf}

\bibitem[Peacock (1968)]{peacock}Peacock, K., 1968.  Multicolor photoelectric photometry of lunar
surface. \emph{Icarus}, \textbf{9}, 16--66

\bibitem[Pearlman, Degnan, \& Bosworth (2002)]{ilrs}Pearlman, M.\,R., Degnan, J.\,J. \& Bosworth, J.\,M., 2002.  The
international laser ranging service. \emph{Adv. Space Res}. \textbf{30},
135--143

\bibitem[Samain et al. (1998)]{oca}Samain, E., and 11 colleagues, 1998.  Millimetric Lunar Laser Ranging at OCA
(Observatoire de la C\^ote d'Azur). \emph{Astron. Astrophys. Suppl. Ser}.
\textbf{130}, 235--244

\bibitem[Shelus (1985)]{mlrs}Shelus, P.\,J., 1985.  MLRS: A lunar/artificial satellite laser
ranging facility at the McDonald Observatory. \emph{IEEE Trans. on
Geoscience and Remote Sensing} \textbf{GE-23}, 385--390

\bibitem[Stubbs, Vondrak, and Farrell (2006)]{stubbs-dust}Stubbs, T.\,J., Vondrak, R.\,R., \& Farrell, W.\,M., 2006.  A dynamic fountain model for lunar dust.  \emph{Advances in Space Research}, \textbf{37}, 59--66

\end{thebibliography}
\end{document}